\documentclass[12pt]{article}
\usepackage{epsfig}
\usepackage{amsfonts}
\usepackage{amsopn}
\usepackage{amsmath}

\title{Dirac equation for strings}

\author{
Maciej Trzetrzelewski \thanks{e-mail: maciej.trzetrzelewski@gmail.com} \\ \\
 M. Smoluchowski Institute of Physics, \\
Jagiellonian University, \\
\L{}ojasiewicza, St. 11, 30-348 Krak\'ow, \\
Poland}

\begin{document}
\date{}
\maketitle

\abstract{Starting with a Nambu-Goto action, a Dirac-like equation can be constructed by taking the square-root of the momentum constraint.
The eigenvalues of the resulting Hamiltonian are real and correspond to masses of the excited string. In particular there are no tachyons. 
A special case of  radial oscillations of a closed string in Minkowski space-time admits exact solutions in terms of wave functions of the harmonic oscillator.}

\section{Introduction}

In 1962 Dirac considered the possibility that leptons may be described by extended objects of spherical topology \cite{Dirac1,Dirac2}. In this way one could understand the muon as an excitation of the ground state of the membrane - the electron. Anticipating Nambu \cite{Nambu} and Goto \cite{Goto} by almost a decade he introduced what is now known as a generalization of the Nambu-Goto action to the case of membranes. Using Bohr-Sommerfeld approximation for the Hamiltonian corresponding to the radial mode, one can then show \cite{Dirac1} that the mass of the first excitation is about $53m_e$ where $m_e$ is the mass of the electron. This is about a quarter of the observed mass of the muon. In fact, when instead of the Bohr-Sommerfeld approximation one uses, more precise, numerical methods one finds that the correct value of the first excitation is about $43m_e$ \cite{gkhk,hadrons}.   Therefore Dirac's model "explains" 1/5th of the actual value of the muon mass. The drawback of that construction was however evident for Dirac from the very beginning - there were no spin matrices involved in the equations, implying that the model could not directly describe fermions.

Recently, the solution to this problem was proposed  \cite{33lewski} by using Dirac's early idea of taking the square-root of constraints. The resulting Dirac-like equation for membranes is a novel way of considering extended objects with spin $1/2$. In particular, the fields appearing there are purely fermionic and (hence) there is no supersymmetry in this approach. The equation is a functional one, however concentrating entirely on radial oscillations one arrives at a relatively simple quantum mechanical problem which can be solved numerically. One then finds that the first excitations is about $61.5m_e$ which is better then $43m_e$ but still not even of the order of the muon mass. At this point, to improve the value, one would have to consider non radial modes. Moreover one could also
relax the assumption that the electromagnetic field is treated as a background i.e. it is not quantized. These  directions are certainly interesting to investigate however they will not be the subject of this paper.

In this paper we study the implications of the above method applied to a string rather then a membrane (we note that a different, than proposed in this paper,  Dirac-like equation for strings was considered already in early 70' \cite{Ramond} - in Section 3 we discuss in more details the difference between that approach and the proposal studied in this paper). There is so far no experimental evidence that elementary particles are extended objects, in particular it is not known whether they are strings, membranes or something else. Hence it is fair to say that string-like objects should also be considered in this framework, even though membranes seem more natural candidates.  It is also clear that the case of strings is equally interesting from the mathematical point of view. We write down the Dirac equation for strings in Minkowski space-time and then study the radial vibrations of the string. This particular case can be solved exactly using the eigen functions of the harmonic oscillator. Introducing a coupling to the electromagnetic field is not possible in this framework, for a very simple reason. To make necessary calculations one needs to consider electromagnetic self-energy of the string, which is well known to be infinite in 3+1 dimensions. Therefore one is forced to consider only neutral strings in this context.

Since the equation considered is Dirac-like it is not a surprise  that the negative eigenvalues of the Hamiltonian appear. However the eigenvalues are, in this model, masses of excited states therefore the appearance of negative masses would be physically unacceptable. In the case of membranes this problem has an elegant solution: the masses of the membrane are measured in units of $\Lambda^{1/3}$ where $\Lambda$ is the membrane world-volume tension. Therefore, negative energies correspond to negative tension since $-\Lambda^{1/3}=(-\Lambda)^{1/3}$. It follows that the Dirac equation for membranes implies that every fermion should have a partner with the same mass, spin and charge but with the negative tension. Due o negative tension these particles would be unstable and therefore not seen in the experiment. However above some critical energy one expects fermions with negative tension to be produced. Let us note that the existence of the $\Lambda \leftrightarrow -\Lambda$ partners, in principle, may have similar implications that we observe in supersymmetric theories - instead of fermion-boson cancellation one expects to see $\Lambda \leftrightarrow -\Lambda$  cancellations of the loop diagrams. This, if true,  would provide a novel approach of the hierarchy problem. Clearly, one cannot make these observations more convincing without a concrete Lagrangian formulation. It is another point which certainly deserves more investigation.

Coming back to  the case of stings. Here   the eigenvalues  are measured in units of $\sqrt{\lambda}$ where $\lambda$ is the string world-volume tension. Consequently the "trick" $\lambda \to -\lambda$ cannot be performed, there are no  $\lambda \leftrightarrow -\lambda$ partners and one  has to assume, following Dirac, that the negative energy/mass states are all filled.

\section{Dirac membrane vs. Dirac string}
In this section we recall certain results of Dirac's membrane model \cite{Dirac1} and compare it to the same results for the string.
The action for a membrane in the presence of the electromagnetic field, considered by Dirac, is a sum of the part corresponding the the electromagnetic field and the membrane part which is the integral of the membrane world-volume.
One can then write down the corresponding equations of motion which for the spherically symmetric case become ($\hbar=c=1$)
\begin{equation} \label{eomr}
\Lambda \frac{d}{dt} \frac{\dot{\rho}}{\sqrt{1-\dot{\rho}^2}}+ \Lambda \frac{2}{\rho\sqrt{1-\dot{\rho}^2}}=\frac{e^2}{8\pi \rho^4},
\end{equation}
where $\rho$ is the radius of the spherical membrane. The l.h.s of (\ref{eomr}) is the part coming from the membrane action while the r.h.s. is $e$ times the electric field on the membrane, $e/2r^2$, per element of the surface.

The repulsive electromagnetic forces and the contractive ones, due to the positive tension $\Lambda$, are balanced when $\dot{\rho}=0$ hence $a^3=e^2/16\pi\Lambda$ where $a$ is the radius of
the electron. On the other hand the total energy of a system at rest $E=e^2/2\rho + \beta \rho^2$   is minimal in the equilibrium provided $\beta=4\pi\Lambda$. That energy is equal to both $3e^2/4a$ and $m_e$ when $\rho=a$. Therefore  we have $a=3e^2/4m_e=0.75r_e$ where $r_e$ is the classical electron radius, $r_e \approx 2.8$fm\footnote{The value $a=2.1$fm is of course not realistic as it is bigger then the charge
radius of the proton $0.87$fm - however this analysis is entirely classical.}.

Using the Hamiltonian formalism for the spherically symmetric case one finds that the Hamiltonian is
\begin{equation} \label{1}
H_r=\sqrt{p_{\rho}^2+(4\pi\Lambda)^2 \rho^4 } +\frac{e^2}{2\rho},
\end{equation}
where $p_{\rho}$ is the momentum corresponding to $\rho$. The quantisation of (\ref{1}) is not clear due presence of the square-root. However one can apply the  the Bohr-Sommerfeld quantization condition
$$
\int_{\rho_{min}}^{\rho_{max}} p_{\rho} d\rho = \pi n , \ \ \  n=1, 2, 3, \ldots
$$
for the $n$th excitation ($n=1$ for muon, $n=2$ for taon). Since $\alpha$ is small one can neglect the potential term in the integral (and hence take $\rho_{min}=0$) which results in
$$
\frac{m_{Dirac}}{m_e} \approx \frac{1}{3}\left(\frac{32\sqrt{\pi} \Gamma(7/4)}{ \alpha
\Gamma(1/4)}n\right)^{\frac{2}{3}}
 = 52.4, \ \ 83.2, \ \ 109.0,   \ldots \ ,$$
where $\alpha$ is the fine structure constant and where we used $4\pi \Lambda = \alpha/4a^3$, $a=3\alpha/4m_e$.  These values are of course not in agreement with the experiment. The mass of the muon is about 4 times larger then $52.4m_e$ while the mass of the taon is off even more.

Let us now apply this technique for the case of a sting. The counterpart of the equation of motion (\ref{eomr}) for the circular sting is
\begin{equation} \label{eomr1}
\lambda \frac{d}{dt} \frac{\dot{\rho}}{\sqrt{1-\dot{\rho}^2}}+ \lambda \frac{1}{\rho\sqrt{1-\dot{\rho}^2}}=e E/2\pi \rho
\end{equation}
where $\lambda$ is the string tension (the factor of $2$ in the second term of the l.h.s. of (\ref{eomr}) becomes $1$).
The r.h.s. is equal to $e$ times electric field on the string divided by $2\pi \rho$ - this quantity is infinite and hence we shall not consider charged strings from now on (i.e. we take r.h.s=0). Classically, neutral strings will collapse to a point after a finite time however quantum mechanically, due to uncertainty principle, a string develops bound states.

To see that explicitly we observe that the Hamiltonian for the spherically symmetric case is now
$$
H_r=\sqrt{p_{\rho}^2+(2\pi\lambda)^2 \rho^2 }
$$
i.e. the square root of the harmonic oscillator\footnote{As expected considering the fact that the spectrum of Nambu-Goto string is given by the spectrum of uncoupled harmonic oscillators.} and hence the spectrum is discrete and the exact eigenvalues are
$$
\frac{m_{Dirac}}{\sqrt{2\pi \lambda}} =  \sqrt{2n+1}, \ \ \ n =0,1, 2, \ldots.
$$

\section{Dirac equation for strings}

Results in previous section are obtained by considering only the radial oscillations of a string.
It is  however well known that when all degrees of freedom are taken into account then one encounters two serious problems: the existence of the tachyon and the violation of Lorentz symmetry. It is frequently argued that supersymmetric extension of bosonic string theory are therefore \emph{necessary} in order to a) avoid the appearance of tachyons from the spectrum and  b) introduce fermions anyway as they exist in the real world.  In this  and the following section we show that a) and b) can be obtained in a different way - not introducing supersymmetry at all.

Let us consider point particles for the moment. The relativistic equation for fermions was obtained by Dirac by taking the square-root of the Klein-Gordon equation, which, in turn,
can be viewed as a constraint derived from the action of a point particle. Therefore the logic behind Dirac equation can be seen as follows
\begin{equation} \label{first}
\hbox{Action}   \rightarrow   \hbox{Constraints}  \rightarrow   \hbox{Klein-Gordon}   \xrightarrow{\sqrt{ \ \ }}  \hbox{Dirac}.
\end{equation}
The last step of this reasoning, the square-root, was originally motivated by Dirac by requiring that the equation should be first order in time derivatives. If we however relax this requirement then
 we could also consider another logic as follows
\begin{equation} \label{second}
\hbox{Action}   \rightarrow   \hbox{Constraints}  \rightarrow   \hbox{Klein-Gordon}  
\end{equation}
$$
 \xrightarrow{\hbox{susy}}  \hbox{massive  Wess-Zumino}.
$$
Both reasonings (\ref{first}) and (\ref{second}) are consistent i.e. they arrive at interesting, albeit different, theories. In fact the supersymmetric massive Wess-Zumino model is mathematically better defined then the Dirac equation since its spectrum is non negative while the spectrum of the Dirac operator is unbounded from below. Therefore, if we were led only by the aesthetic arguments e.g.  "nice mathematics",  we could in fact claim that (\ref{second}) is  more superior. However at the same time (\ref{second}) was never found in Nature while (\ref{first}) was.

Let us come back to the case of extended objects.  Considering the above remarks it would be certainly incomplete to consider only the analog of (\ref{second})  (whose analysis can be found in every string theory text book) and completely forget about (\ref{first}). After all (\ref{first}) works (i.e. is supported experimentally) while (\ref{second}) has never been experimentally confirmed - one should therefore investigate (\ref{first}) as well, in the case of extended objects.

In \cite{33lewski}, we used this point and generalized (\ref{first}) to the case of membranes in flat space, however clearly the same can be done for the string.
Let us start with the Nambu-Goto action
\begin{equation}
S=-\lambda\int\sqrt{-G}d\tau d\sigma,
\end{equation}
$$
G=\det G_{\alpha \beta}, \ \ \  G_{\alpha\beta}=\partial_\alpha X^\mu \partial_\beta X^\nu \eta_{\mu\nu}
$$
where $X^{\mu}$, $\mu=0,1,\ldots, D-1$ are the coordinates parametrized by $\sigma^{\alpha}$'s, $\alpha=0,1$ ($\sigma^0=\tau, \sigma^1=\sigma$), $\lambda$ is the tension, $\eta_{\mu\nu}$ is the Miknowski metric - we use the many-minus convention throughout the paper. It follows that the canonical momenta $\mathcal{P}_\mu:=\delta S/\delta\partial_0X^\mu$ satisfy the constraints
\begin{equation} \label{constraint}
\eta^{\mu\nu}\mathcal{P}_{\mu}\mathcal{P}_{\nu}+\lambda^2 G_{11}=0,
\end{equation}
and
\begin{equation} \label{constraint1}
\partial_{\sigma}X^{\mu}\mathcal{P}_{\mu}=0.
\end{equation}
Constraint (\ref{constraint}) is an analog of the mass-shell constraint for the point particle therefore we would like to take the square-root of that equation. It can be done in at least two different ways, one is to write
\begin{equation} \label{1waya}
\left( -i\gamma^{\mu} \frac{\delta}{\delta X^{\mu}}+ \lambda i \gamma^{\mu}\partial_{\sigma}X_{\mu}   \right)\Psi=0
\end{equation}
where we substituted the canonical momenta with the functional derivative $\mathcal{P}_{\mu} \to -i\delta/\delta X^{\mu}$. The spinor functional $\Psi$ was introduced at the same time (the domain of  $\Psi$ is a space of all possible configurations of the string). Note that in the case of a membrane the counterpart of the term  $\gamma^{\mu}\partial_{\sigma}X_{\mu} $ involves the Poisson bracket, $\gamma^{\mu\nu}\{X_{\mu},X_{\nu}\}$, while for a general object with $p>2$ spatial dimensions we have the Nambu bracket.

Second way is to leave the square-root explicitly i.e. consider
\begin{equation} \label{2way}
\left( -i\gamma^{\mu} \frac{\delta}{\delta X^{\mu}}+ \lambda\sqrt{ -\partial_{\sigma}X_{\mu}\partial_{\sigma}X^{\mu} } 1  \right)\Psi=0.
\end{equation}

Both equations, (\ref{1waya}) and (\ref{2way}), are good candidates for the Dirac equation for the string. They are however very different in terms of their spectrum. In the next section we will argue that, when concentrating entirely on the radial excitations, the spectrum of (\ref{1waya}) is continuous while the spectrum of (\ref{2way}) is discrete. Finlay let us note that both (\ref{1waya}) and (\ref{2way}) define a Hamiltonian operator via
\begin{equation} \label{totalhamiltonian}
H= \int d\sigma \mathcal{H}, 
\end{equation}
$$
\mathcal{H}:= i\frac{\delta}{ \delta X^{0}} = \begin{cases} \gamma^0(-i\gamma^{k} \frac{\delta}{\delta X^{k}}+\lambda  \gamma^k \partial_{\sigma}X_k  ), &  \\  \gamma^0(-i\gamma^{k} \frac{\delta}{\delta X^{k}}+ \lambda\sqrt{ -\partial_{\sigma}X_{k}\partial_{\sigma}X^{k} } 1    ), &  \end{cases}
$$
($k=1,2,3$), when working in the temporal gauge $X^0=\tau$.
\\

In early 70' a certain Dirac-like equation for strings was also considered \cite{Ramond}, although motivated differently.
It was observed that one way to generalize the Dirac equation, into the case of a string, is to consider $<\Gamma^{\mu}P_{\mu}>_0+ \ m=0$  where $<A>_0$ is understood as the $\tau$-zero-mode part of the operator $A$, $P_{\mu}$ is the momentum operator for the string, $\Gamma^{\mu}$ are the generalized, $\sigma$ and $\tau$ dependent, Dirac matrices.  

That construction still contains $m$ as a parameter, moreover equation $<\Gamma^{\mu}P_{\mu}>_0+ \ m=0$ is not a square-root of the quadratic momentum constraint for a string (\ref{constraint}) (not just because $m$ is not present in (\ref{constraint}) but also because the introduction of $\sigma$ and $\tau$ dependent Dirac matrices $\Gamma^{\mu}$ is not needed when taking the square-root of (\ref{constraint})). Therefore it is a completely different approach and a different equation compared to our proposal (\ref{1waya}) or (\ref{2way}).

\section{$SO(2)$ symmetric string}

Equations (\ref{1waya}) and (\ref{2way}) are obtained by introducing the wave-functional $\Psi$ and demanding that the square-root of the constraint, augmented to the functional operator form, is satisfied on $\Psi$. This approach seems difficult if the shape of the string is arbitrary - due to the functional character of the equation. On the other hand if one considers $SO(2)$ symmetric case one arrives at a relatively simple quantum mechanical system described by only one degree of freedom (the radial variable) instead of infinitely many of them.

This approach is similar to what one does in the mini-superspace approaches in quantum gravity pioneered by DeWitt \cite{DeWitt}. There, the quantum version of the constraint (the Hamiltonian constraint) is the Wheeler-DeWitt equation - which is a functional one.  This equation simplifies considerably if one concentrates entirely on the metrics of the  Friedmann-Robertson-Walker type where there is only one degree of freedom - the scale factor. 

Even better example can be found in the case of the Schr\"odinger equation for Yang-Mills theories in the temporal gauge \cite{Greensite1,Greensite2,Greensite3}. There, the hamiltonian, in structure, is similar to (\ref{totalhamiltonian}) (unlike for gravity where $H=0$). Of course there are significant differences e.g. (\ref{totalhamiltonian}) is 1st order in functional derivatives and contains gamma matrices, unlike the hamiltonian for gauge theories.

Let us consider the case of $SO(2)$ symmetric strings in $3+1$ dimensional Minkowski space. The coordinates of such string can be always chosen so that
$$
X^0=\tau, \ \ \ X^1=\rho(\tau) \cos \sigma, \ \ \ X^2=\rho(\tau) \sin \sigma, \ \ \ X^3=0
$$
i.e. the string is laying in the $X^1X^2$ plane. Let us also introduce the averaged (over the volume) momentum operator
$$
p_{\mu} := \int_{S^1} d\sigma \mathcal{P}_\mu
$$
at some time-slice. As a result, equation (\ref{2way}) gives
\begin{equation} \label{radialeq}
(-i\gamma^\mu \partial_{\mu}|_{\rho}+2\pi \lambda \rho )\psi=0
\end{equation}
where we replaced the momenta $p_{\mu}\to -i\partial_{\mu}$ by ordinary derivatives;  the symbol $|_{\rho}$ means taking only the radial part  of the operator.

Note that the analogous equation for (\ref{1waya}) is like in (\ref{radialeq}) but without the potential term i.e. it is $-i\gamma^\mu \partial_{\mu}|_{\rho}\psi$=0. This is due to the fact that the integral over the potential term vanishes which is true also for the complete Hamiltonian (\ref{totalhamiltonian}), i.e. for an arbitrary closed string, due to periodic boundary conditions $X^k(0)=X^k(2\pi)$, $k=1,2,3$. Therefore its spectrum is continuous and we shall not be considering this case in the remaining part of this paper.

On the other hand equation (\ref{radialeq}) has  discrete family of solutions. To see this let us write the operator $-i\partial_{\mu}$ in new coordinates $(\tau,\rho,\sigma,z)$. We have
$$
-i\gamma^{\mu}\partial_{\mu}\psi =\left(-i\gamma^0\partial_{\tau}-i\gamma^1 e^{-\sigma \gamma^1\gamma^2}\partial_{\rho}-i\frac{1}{\rho}\gamma^2 e^{-\sigma \gamma^1\gamma^2}\partial_{\sigma}-i\gamma^3 \partial_z \right)\psi.
$$
Our objective now is to single out the radial part of the above operator. To do so we rotate the spinor $\psi$ and define $\phi:= e^{-\frac{\gamma^1\gamma^2}{2}\sigma }\psi$. We now find that
$$
-i\gamma^{\mu}\partial_{\mu}\psi =e^{\frac{\gamma^1\gamma^2}{2}\sigma}\mathcal{D}\phi,
$$
$$
\mathcal{D}:= -i\gamma^0 \partial_{\tau}-i\gamma^1\left(\partial_{\rho}+\frac{1}{2\rho}\right)-i\gamma^2 \frac{1}{\rho}\partial_{\sigma}-i\gamma^3 \partial_z     
$$
therefore equation (\ref{radialeq})  becomes
\begin{equation} \label{radialeq1}
\mathcal{D}\phi +2\pi\lambda \rho\phi =0.
\end{equation}
In order to solve (\ref{radialeq1}) let use use another ansatz $\phi=\rho^{-1/2}e^{-iE \tau+im\sigma}\chi(\rho)$ where $E$ is the energy (mass of the excited state) and $m \in \mathbb{Z}$ (note that we are assuming no $z$ dependence). Using the Dirac representation of the gamma matrices and splitting the spinor $\chi=(\chi_1,\chi_2,\chi_3,\chi_4)^T$ by defining
$$
\chi_+ = (\chi_1, \chi_4)^T, \ \ \ \ \chi_{-} = (\chi_2, \chi_3)^T
$$
we find from   (\ref{radialeq1}), that $\chi_{\pm}$ decouple and satisfy
\begin{equation} \label{eigeneqn}
H_{\pm}\chi_{\pm}= \epsilon \chi_{\pm},  \ \ \ \
H_{\pm}=\left(\begin{matrix} 
  x & i\partial_x \pm \frac{i m}{x} \cr
 i\partial_x \mp \frac{i m}{x} & - x
\end{matrix}\right),
\end{equation}
$$
x=\rho \sqrt{2\pi \lambda }, \ \ \ \  \epsilon =E/\sqrt{2\pi \lambda}
$$
where $x$ and $\epsilon$ are dimensionless.

In general, the spectrum of $H$ can be found using numerical methods however the case $m=0$  is much easier and can be solved exactly.
To see this explicitly let us write $\chi_{\pm}$ as  $\chi_{\pm}=(F, i G)^T$ and define  $a_{\pm}=F \pm G$. The eigen equation (\ref{eigeneqn}) implies that
$$
(x \pm \partial_x)a_{\mp} = \epsilon a_{\pm} \ \ \ \to \ \ \ (-\partial_x^2 + x^2)a_{\pm}= (\epsilon^2\mp1)a_{\pm}.
$$
Therefore the components $a_{\pm}$ decouple and are given by the solutions of the harmonic oscillator
$$
a_{+ \ n}=\psi_n^{H.O.}, \ \ \ \ a_{-\ n}=\psi_{n+1}^{H.O.},  \ \ n=0,1,2,\ldots \ ,
$$
where $\psi_n^{H.O.}$ is the $n$th eigenstate of the harmonic oscillator. They corresponding to energy-squared $\epsilon^2=2(n+1)$.  It is instructive to substitute them
directly into (\ref{eigeneqn}) - we will find that they correspond to positive energy $\sqrt{2(n+1)}$. However there exist another family of solutions given by
$$
a_{+ \ n}=\psi_n^{H.O.}, \ \ \ \ a_{-\ n}=-\psi_{n+1}^{H.O.},  \ \ n=0,1,2,\ldots \ \ .
$$
These solutions correspond to negative masses $\epsilon=-\sqrt{2(n+1)}$ and therefore should be eliminated somehow from the theory. Because they are fermion states one could assume that these states are already occupied i.e. introduce the Dirac sea of states.

The case of arbitrary $m$ also contains real and discrete spectrum. Reality is  guaranteed by the fact that   $H_{\pm}$ is hermitian.
The proof that $H_{\pm}$ is discrete can be obtained by noting that the square of $H_{\pm}$ is
\begin{equation}  \label{discr}
H_{\pm}^2= H_{\pm, m=0}^2
+ \left( \begin{array}{cc}
                    \frac{m(m\mp1)}{x^2}  & 0    \\
                      0  &  \frac{m(m\pm1)}{x^2}
                              \end{array} \right).
\end{equation}
The second operator on the r.h.s. of (\ref{discr}) is positive for any $m$ while the first is discrete therefore $H_{\pm}^2$ is bounded from below by a discrete operator. By a standard theorem in functional analysis \cite{func} this implies that $H_{\pm}^2$, and hence $H_{\pm}$, is discrete.

\section{Summary}

Dirac equation is one of the most successful, moreover extremely simple and elegant, equations describing elementary particles.
The logic behind that equation is based on two assumptions: the equation should be Lorentz covariant and first order in time derivatives. The later rules out the Klein-Gordon equation (which is also known to fail for other reasons) while the first rules out the Schr\"odinger equation. A correct solution of the problem turns out to be given by the square-root of the Klein-Gordon equation or equivalently the square-root  of the quadratic (in momenta) constraints of the classical theory -  which are then replaced by the corresponding differential operators. Because this heuristics works one is tempted to apply this procedure also to extended objects.

In the original Dirac equation for a point particle the mass is a parameter of the theory that one has to fix using the experimental data. It is not determined by any equation. When considering Dirac equation for extended objects one arrives at functional differential operator whose eigenvalues are masses of the object. If that spectrum is discrete then there exists a possibility that one could explain lepton  masses as excitations of the ground state. This idea for the first time was proposed by Dirac \cite{Dirac1,Dirac2} in the case of a bosonic membrane.

In this paper we use this idea concentrating on the case of Dirac equation for a string rather then a membrane. Because in 3+1 dimensions, a charged string introduces infinite electromagnetic potential on the string, one is forced to consider neutral strings only.  There are at least two different ways (equations (\ref{1waya}) and (\ref{2way})) in which one can perform the square-root of the constraint.  Considering only the radial excitations we argued that the spectrum of (\ref{1waya}) is continuous while the spectrum of (\ref{2way}) is discrete. The analysis is only approximate as we are neglecting all other degrees of freedom however, we claim that these conclusions are the same in the general case.

An important property of this model is that the spectrum of the resulting Hamiltonian is given by masses of excited states of the string. The Hamiltonian is hermitian hence all the masses are real and therefore there is no problem with tachyon excitation, as it is the case in bosonic string theory. Negative mass states appear and cannot be interpreted as strings with negative tension (as it is the case for membranes). Therefore one needs to assume that these states are all filled by introducing the Dirac sea of states with negative masses. This may be seen as a serious drawback of the model.

\section{Acknowledgements}
This work was supported by DFG (German Science Foundation) via the SFB grant.

\end{document}